\documentclass[prc,aps,nofootinbib,showkeys,showpacs,twocolumn]{revtex4}
\usepackage{epsfig}
\usepackage{graphicx}
\usepackage{amssymb}
\usepackage{color}
\usepackage{upgreek}
\usepackage{mathtools}
\usepackage{comment} 
\usepackage{tcolorbox}

\usepackage{braket}
\usepackage{comment}
\usepackage{ulem}

\begin{document}

\title{Three-flavor neutrino oscillations using the Phase Space Approach}

\author{Mariane Mangin-Brinet} \email{mariane@lpsc.in2p3.fr}
\affiliation{Laboratoire de Physique Subatomique et de Cosmologie, CNRS/IN2P3, 38026 Grenoble, France}

\author{Angel Bauge} 
\affiliation{Universit\'e Paris-Saclay, CNRS/IN2P3, IJCLab, 91405 Orsay, France}

\author{Denis Lacroix } \email{lacroix@ijclab.in2p3.fr}
\affiliation{Universit\'e Paris-Saclay, CNRS/IN2P3, IJCLab, 91405 Orsay, France}

\date{\today}
\begin{abstract}
The Phase-Space Approximation (PSA) approach, originally applied in [Phys. Rev. D 106, 123006 (2022)] to describe neutrino oscillations from a stellar object in the two-flavor limit, is extended here to describe the more realistic case where neutrinos can
oscillate between three different flavors. The approach is
successfully validated against the exact solutions up to eight neutrinos. In all cases where the exact solution is feasible, the PSA provides excellent reproduction of the neutrino oscillation dynamics. By replacing the full problem with a set of simple mean-field equations, the PSA offers a versatile, predictive, and easily parallelizable approach for tackling three-flavor problems. This enables the simulation of large-scale neutrino oscillations, as illustrated here with simulations involving up to 300 neutrinos. Additionally, the method provides insight into the system's equilibration properties.
\end{abstract}



\maketitle

\section{Introduction}

Understanding the oscillations of neutrinos emitted from stellar objects is a complex problem that involves numerous interrelated physical processes. These include their creation within stellar objects, interactions with electrons, and direct neutrino–neutrino interactions occurring shortly after emission \cite{Ful87,Not88,Sig93,Dua06,Bah07,Vol24}. To gain insight into these dynamics, simplified models have been proposed and used to study how such interactions influence neutrino oscillations, including the possible entanglement between different neutrino beams \cite{Peh11,Bir18,Pat19,Rra19,Cer19,Pat21,Mar21,Rog21,Cer22,Bal22,Xio22,Lac22,Rog22a,Mar23a,Mar23b,Bha23,Lac24}. These models have revealed a variety of intriguing phenomena, such as beam-splitting processes, neutrino entanglement, equilibration arising from mutual neutrino interactions, and decoherence effects induced by interactions between neutrino beams and their surrounding environments.

With the recent rise of quantum computing, simplified Hamiltonian simulations of neutrino oscillations in flavor space have attracted considerable attention. In recent years, several research groups have investigated the potential of using the neutrino oscillation problem as a pilot application for current and future quantum computing platforms, either within the simplified two-flavor framework \cite{Hal21,Yet22,Kum22,Ill22a,Ill22b,Ami23,Siw23b}, or in the more realistic three-flavor case \cite{Tur24,Che25,Spa25}. Due to the limitations of present-day quantum hardware, simulations have so far been restricted to relatively small numbers of neutrinos and short evolution times. Initial steps have also been taken toward understanding the quantum information aspects of the three-flavor case \cite{Siw23,Siw25}, as well as exploring the classical simulability of the problem through the quantification of quantum "magic" \cite{Che25b}.

These developments have, in parallel, motivated the exploration of classical computational methods aimed at pushing the limits of how many neutrinos, including their mutual interactions, can be accurately simulated. Exact solutions are typically feasible only for relatively small systems: up to about 20 neutrinos in the two-flavor approximation and no more than 10 in the three-flavor case. 
Tensor network methods have also been explored, though so far only for two-flavor systems \cite{Rog21, Cer22}. These methods allow simulations of a few tens of neutrinos, provided that the entanglement entropy does not grow too fast. However, tensor networks such as matrix product states (MPS), while viable in some contexts, are not ideally suited for systems with all-to-all interactions, such as those between neutrinos in different beams. An alternative approach, called Phase-Space Approximation (PSA), has been recently applied to the neutrino oscillation problem. This method has a long history in low-energy nuclear physics and has been successfully applied to systems with all-to-all interactions \cite{Ayi08,Lac12,Lac13,Lac14,Yil14,Lac14b,Lac16,Reg18,Ulg19,Czu20}. 
By mapping a quantum many-body problem into a statistical set of "simple" trajectories propagated in phase-space, PSA offers an approximate yet scalable approach capable of handling systems composed of large numbers of particles.
The PSA method was first adapted to the neutrino oscillation problem in the case of two interacting neutrino beams \cite{Lac22}, demonstrating strong predictive power, including the ability to capture entanglement properties.  In this context, a "beam of neutrinos" refers to a set of neutrinos that share similar properties. The approach was subsequently extended to an arbitrary number of beams, still within the two-flavor approximation \cite{Lac24}. In particular, it was shown that it can simulate several hundred neutrinos on a standard personal computer. 

We believe that the development of such an approach is crucial for several reasons. First, by enabling simulations of systems ranging from very few to many neutrinos, it offers the potential to explore and to understand the transition from the microscopic to the macroscopic regime -- a key aspect in the context of neutrino physics. Second, one can see how physical phenomena related to the many-body problem, such as entanglement, scale with the number of particles. 
Third, assuming that quantum computers continue to advance and eventually support the accurate simulation of systems with large numbers of qubits or particles, the PSA approach could serve as a valuable classical benchmark for comparison. 
Finally, demonstrating the efficiency of PSA, which is inherently a classical computational method, might be a way to explore in practice the limit of classical simulability for complex quantum systems.

The main objective of the present work is to extend the PSA method
to the treatment of three-flavor neutrino oscillations. After recalling
the Hamiltonian commonly used on quantum computers to model the three-flavor oscillation problem in the presence of neutrino-neutrino interactions, we describe in detail how the PSA framework can be adapted to this more complex problem. We then benchmark it against the exact solution when available. Finally, we demonstrate the scalability of the approach by presenting simulations involving several hundred neutrinos.

\section{Hamiltonian, notations and parameters}

We start with a system of $N$ neutrinos having three possible flavors $(e, \nu, \tau)$. We assume that their evolution is governed by the following Hamiltonian \cite{Siw23,Tur24,Che25,Spa25,Siw25,Che25b} expressed in the mass basis:  
\begin{eqnarray}
H=H_{\nu}+H_{\nu\nu} = \sum_{\alpha=1}^{N} \vec{B}\cdot \vec{Q}(\alpha)+ \sum_{\alpha \neq \beta}^{N} \mu_{\alpha\beta}\vec{Q}(\alpha)\cdot\vec{Q}(\beta). \label{eq:Hneut}
\end{eqnarray}
The term $H_\nu$ is a one-body term describing vacuum oscillations, while $H_{\nu\nu}$
is the two-body term describing coherent neutrino-neutrino interaction. 
Indices $\alpha,\beta$ label the neutrinos, $\mu_{\alpha\beta}$ is the coupling strength between the neutrinos, and the set of operators $\vec{Q}$ are given by:
\begin{eqnarray}
    Q_{m}(\alpha)&=&\frac{1}{2}\sum_{i,j=1}^{3}a^{\dagger}_{i}(\alpha)(\lambda_{m})_{ij}a_{j}(\alpha) \label{eq:Q_def}
\end{eqnarray}
where $\lambda_m$ are the Gell-Mann matrices ($m=1, \ldots,8$). $(a^{\dagger}_{i}(\alpha), a_{i}(\alpha))$ operators are the creation and annihilation operators of neutrino $\alpha$ in the mass basis. Since we consider three-flavor neutrinos, $i=1,2,3$.
There exist many conventions and sets of parameters for the Hamiltonian, namely the choice of $\vec{B}$ and $\mu_{\alpha\beta}$. We will stick to one set of parameters, those used in \cite{Tur24}, 
with which we will compare our results. In this reference, the coupling between neutrinos is assumed to be time-independent.  
Nevertheless, it should be stressed that our method does not depend on the specific Hamiltonian considered but can be applied to any Hamiltonian, including time-dependent ones (see, for instance, Ref. \cite{Lac22}, where the PSA approach is applied for a time-dependent Hamiltonian for the two-flavor application case). This specificity is a clear asset of the PSA method. 

For the sake of clarity, we briefly recall the notations and parameters used in \cite{Tur24} below. 
The vacuum interaction is given by 
\begin{eqnarray}
H_{\nu}=\frac{\mu}{N}\sum_{\alpha=1}^N  \Big(- \frac{\omega}{2\Omega} \lambda^{\alpha}_3  + \frac{\omega-2\Omega}{2\sqrt{3}\Omega} \lambda^{\alpha}_8 \Big),
 \label{eq:Hnu}\end{eqnarray}
where the oscillation frequencies are defined by:
\begin{eqnarray*}
\omega=\frac{\Delta m_{21}^2}{2E}, \qquad \Omega=\frac{\Delta m_{31}^2}{\sqrt{3}E},
\end{eqnarray*}
 where $\Delta m_{ij}=m^2_i-m^2_j$ is the squared mass difference between two mass-eigenstate neutrinos. 
 Note that all neutrinos have the same energy $E$. 
The neutrino-neutrino interaction term is:
\begin{eqnarray*}
H_{\nu\nu}=\frac{\mu}{N}\sum_{\alpha<\beta}^{N}  (1-\cos{\theta_{\alpha\beta}})\vec{\lambda}_{\alpha}\cdot  \vec{\lambda}_{\beta} . \label{eq:2body-term}
\end{eqnarray*}
We will use for $\vec{B}$ and $\mu_{\alpha\beta}$ the following values (in mass basis):
\begin{eqnarray*}
\vec{B}^T&=& \left(0,0,- \frac{\omega}{\Omega},0,0,0, -\frac{2\Omega-\omega}{\sqrt{3}\Omega}\right), \\
\mu_{\alpha\beta}&=&2\frac{\mu }{N}(1-\cos{\theta_{\alpha\beta}}),
\end{eqnarray*}
with 
$\displaystyle \theta_{\alpha\beta}=\arccos(0.9)\frac{|\alpha-\beta|}{N-1}$, and where the neutrino coupling
constant $\mu$ has been defined such that the one-body and two-body terms have the same magnitude. Following \cite{Tur24}, we will display the results in unit of this coupling constant $\mu$. 

In the following, we will always start from an initial state given in the flavor basis, to study the oscillation of neutrinos from one flavor to the others. We then have two possible choices: either we rotate the mass-basis Hamiltonian and perform the evolution in the flavor basis, or we rotate the initial
state from the flavor basis to the mass basis, perform the evolution in the mass basis, and rotate back the propagated solution to the flavor basis.  
In the former case, the two-body interaction is left invariant under the basis change, and we
only need to consider the effect of the rotation on the vacuum interaction term, Eq.  (\ref{eq:Hnu}). If we write the one-body Hamiltonian in the mass basis as $H_{\nu}= \sum_{\alpha} H^{(\alpha)}_{\nu}$, each one-neutrino term $H^{(\alpha)}_{\nu}$
is transformed as 
\begin{eqnarray*}
H^{(\alpha), {\rm fl}}_{\nu}=U_{\rm PMNS}H^{(\alpha)}_{\nu} U^{\dagger}_{\rm PMNS},
\end{eqnarray*}
where $U_{\rm PMNS}$ is the Pontecorvo-Maki-Nakagawa-Sakata (PMNS) matrix, given by:
\begin{widetext}
\begin{eqnarray*}
U_{\rm PMNS}={\small{
\begin{pmatrix}
c_{13} c_{12} & c_{13} s_{12} & s_{13}e^{-i\delta} \\
-s_{12} c_{23}-c_{12} s_{23}s_{13} e^{i\delta}   &  c_{12} c_{23}-s_{12} s_{23}s_{13} e^{i\delta} & s_{23}c_{13} \\
s_{12} s_{23}-c_{12} c_{23}s_{13} e^{i\delta}   &   -c_{12} s_{23}-s_{12} c_{23}s_{13} e^{i\delta} & c_{23}c_{13} 
\end{pmatrix}
}}.
\end{eqnarray*}
\end{widetext}
Notations $c_{ij}$ and $s_{ij}$ are used to denote $\cos\theta_{ij}$ and $\sin\theta_{ij}$ respectively, and $\delta$ is the CP violation phase. 
If the evolution is performed in the mass basis, then the Hamiltonian is directly given
by Eq. (\ref{eq:Hneut}), but the initial state is rotated from the flavor basis to the
mass basis using the PMNS matrix, and the final state undergoes reverse rotation.

The parameters we use in the following calculations are taken from Refs. \cite{Est20, Nufit24}, assuming normal ordering, and are the same as those used in \cite{Tur24}. For the sake of completeness, these parameters are given in Table \ref{Tab:Neut_Param}. 
\begin{center}
\begin{table}
\begin{tabular}{|c|c|c|}\hline
Parameter &  Values  in \cite{Tur24}\\\hline
$\delta m^{2}$ &   7.41 $\times 10^{-17}$ MeV$^2$ \\
$\Delta m^{2}$ & 2.505 $\times 10^{-15}$ MeV$^2$ \\
$\theta_{12}$ & 33.67$^o$ \\
$\theta_{13}$ & 8.58$^o$\\
$\theta_{23}$ & 42.3$^o$ \\
$\delta$ &  232$^o$ \\\hline
\end{tabular}
\caption{Parameter values used in our simulation (taken from Refs. \cite{Est20, Nufit24}).}\label{Tab:Neut_Param}
\end{table}
\end{center}

Due to the rapidly increasing size of the Hilbert space ($\Omega = 3^N$), 
an exact solution of the problem is feasible only for a relatively small number of neutrinos ($N \le 10$). 
Such exact solutions can be obtained simply by directly diagonalizing the Hamiltonian in the full Hilbert space. In the present work, we developed such an exact treatment, which we validated by comparing to 
the results reported in \cite{Tur24}. Wherever exact solutions are available, they will be used to assess the predictive power of the PSA approach for the three-flavor neutrino oscillation problem.

\section{Phase-Space Approach for the three-flavor neutrino oscillations}

The Phase-Space Approximation, originally developed in the context of nuclear physics problems \cite{Ayi08,Lac12} (see also \cite{Lac14} for a review) maps
the complex evolution of interacting particles onto a statistical ensemble of independent, simpler trajectories, each initialized with stochastically sampled initial conditions.

This approach was originally introduced to overcome the limitations of mean-field theory, which, while often accurate in predicting the expectation values of one-body observables, typically fails to describe the quantum fluctuations of these observables stemming from genuine many-body correlations beyond the independent quasi-particle picture.  

The approach relies on two key ingredients: (i) the strategy used for the initial sampling, and (ii) the choice of the "simple" trajectories. While the 
probability distribution to sample initial condition is not unique \cite{Yil14,Ulg19,Lac24}, the sampling strategy we used is designed to ensure that the statistical average over the ensemble of initial conditions accurately
reproduces the quantum fluctuations, with at least the requirement that the first and second moments of the one-body observables are correctly reproduced. As for the trajectory, the simplest and most commonly adopted choice, also employed in the present work, is to evolve each trajectory according to the mean-field equations of motion.

Previous applications of the PSA to neutrino systems have been limited to the two-flavor approximation, in which the problem can be formulated within the SU(2) framework. In this case, the PSA treatment closely resembles its earlier use in models such as the Lipkin model \cite{Lac12,Yil14,Ulg19}. A central objective of the present work is to extend the PSA approach to the more realistic three-flavor scenario, corresponding to the SU(3) case. The equations of motion and the initial sampling technique are presented below, followed by an illustration of the approach.

\subsection{Mean-field Equations of motion}
\begin{figure*}[htbp]
\begin{center}
\includegraphics[height=6.8cm,width=8cm]{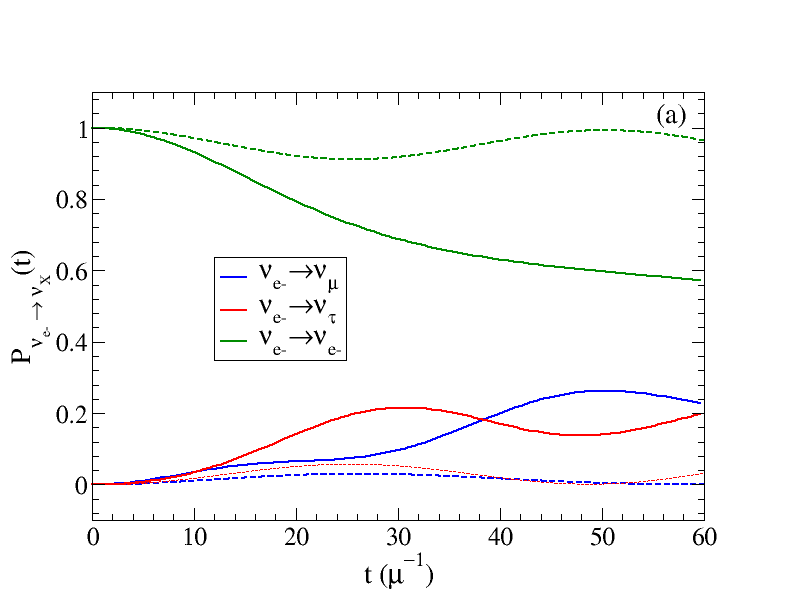}  
\hspace{-0.5cm}
\includegraphics[height=6.8cm,width=8cm]{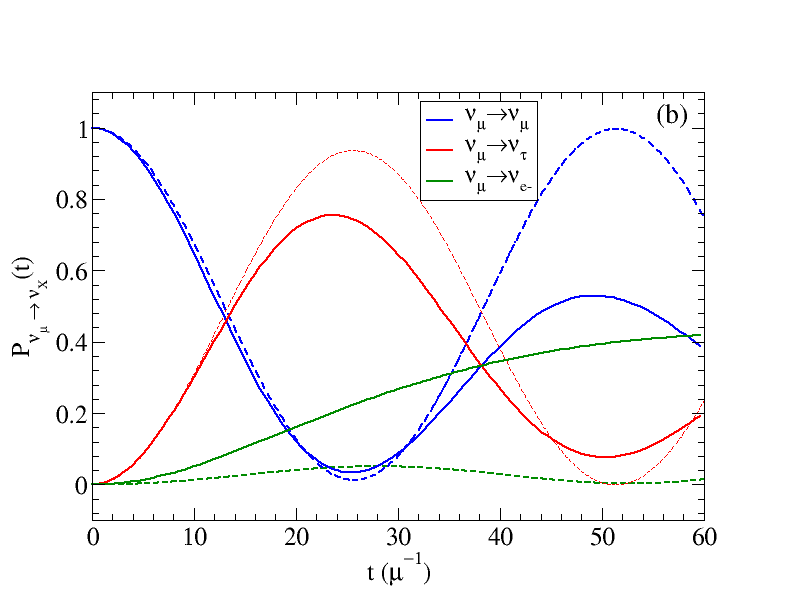}  
\includegraphics[height=6.8cm,width=8cm]{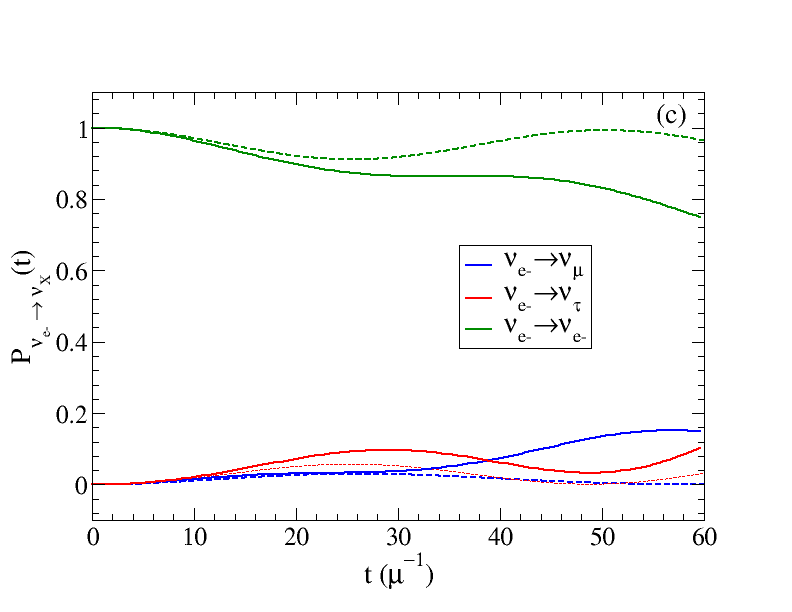}  
\hspace{-0.5cm}
\includegraphics[height=6.8cm,width=8cm]{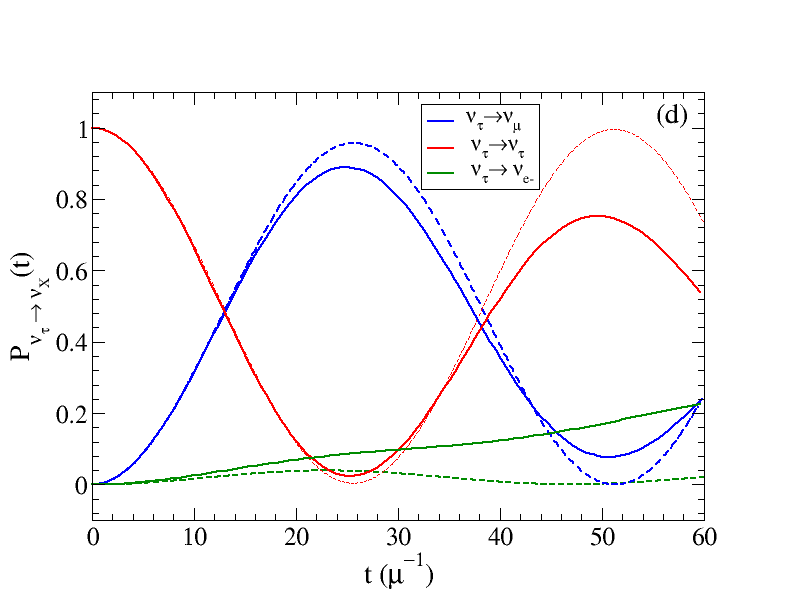}  
\caption{{\it{Comparison of flavor evolutions of 8 interacting neutrinos, initialized to $| \nu_e\nu_{\mu}\nu_e \nu_{\tau} \nu_{\tau}\nu_e\nu_{\mu}\nu_{e} \rangle$ (from left to right and top to bottom). Since the initial state is symmetric, only the first four l.h.s. neutrinos are represented. The flavor evolution is computed exactly (solid lines) and using a mean-field approximation (dashed lines). 
Mean-field solutions fail to describe the exact flavor evolution of neutrinos, even for short times. 
}}}\label{fig:mfvsexact}
\end{center}
\end{figure*}

In its standard formulation, the equations of motion used along each trajectory of the Phase Space approach are simply considered to be the mean-field equations of motion. These equations are rather standard and well established in the literature \cite{Das08,Siw25}. However, 
to avoid any confusion due to the variety of notations commonly used, and although this derivation is relatively standard, 
we find it useful to recall here some of the key steps to obtain these equations of motion using the Ehrenfest theorem for the set of one-body operators $Q_m$. 

The operators $\vec{Q}(\alpha)$ defined by Eq. (\ref{eq:Q_def}) satisfy the following commutation relations:
\begin{eqnarray*}
[Q_m(\alpha),Q_{m'}(\beta)]&=&i\delta_{\alpha\beta} \sum_{c=1}^{8}f^{mm'c}Q_c(\alpha),
\end{eqnarray*}
where $f^{mm'c}$ are the SU(3) structure constants. The Ehrenfest theorem applied to operators $\vec{Q}$ reads 
\begin{eqnarray*}
i  \frac{d}{dt} \langle Q_{m}(\alpha) \rangle &=& \langle [Q_{m}(\alpha), H] \rangle .
\end{eqnarray*}
Using the previous commutators and the mean-field approximation:
\begin{eqnarray}
\langle Q_{m}(\alpha) Q_{m'}(\beta)\rangle \simeq \langle Q_{m}(\alpha)\rangle\langle Q_{m'}(\beta)\rangle,
\label{eq:nofluc}
\end{eqnarray}
we obtain the following equations of motion (where we have also taken into account the fact that the coupling constant $\mu_{\alpha\beta}$ is symmetric):
\begin{eqnarray*}
&&\frac{d}{dt} \langle Q_{m}(\alpha) \rangle = 
  \sum_{c=1}^{8}\sum_{m'}B_{m'} f^{mm'c} \langle Q_c(\alpha) \rangle 
 + 2 \sum_{ \beta\ne \alpha} \mu_{\alpha\beta}  \\
&& \times \sum_{c, m'<c}^{8}f^{mm'c}\left(   \langle Q_c(\alpha) \rangle  \langle Q_{m'}(\beta)  \rangle- \langle Q_{m'}(\alpha) \rangle  \langle Q_{c}(\beta)  \rangle \right).
\end{eqnarray*}

Defining a generalized cross product of two 8-dimensional vectors as $(\vec{A}\times \vec{B})_a=\sum_{b,c}f^{abc}A_bB_c$, the previous equations take the compact form:
\begin{eqnarray*}
\frac{d}{dt} \vec{\langle Q \rangle}(\alpha)  &=& \vec{B}\times \vec{\langle Q \rangle}(\alpha) 
+\sum_{ \beta\ne \alpha} \mu_{\alpha\beta}\vec{\langle Q \rangle}(\beta) \times  \vec{\langle Q \rangle}(\alpha) .
\end{eqnarray*}
Note that here we restrict the application to one neutrino per neutino beam. However, having more neutrino per beam in the mean-field does not increase the numerical cost in mean-field. The PSA method presented below inherited also this from the mean-field.   
Finally, since we will focus on determining the neutrino polarization, we substitute the operators $\vec{Q}(\alpha)$ by the (eight-component) polarization defined as $\vec{P}(\alpha)=2\langle \vec{Q}(\alpha)\rangle$, and the equations of motion
finally read:
\begin{eqnarray}
\frac{d}{dt} \vec{ P }(\alpha)  &=& \vec{B}\times \vec{ P}(\alpha) 
+\frac{1}{2}\sum_{ \beta\ne \alpha} \mu_{\alpha\beta}\vec{ P }(\beta) \times  \vec{ P}(\alpha).  \label{Eq:EqOfMotion}
\end{eqnarray}
For $N$ neutrinos, the mean-field evolution reduces to a set of $8N$ non-linear 
coupled equation that can easily be solved even for large $N$, using, for instance, the Runge-Kutta 
time-integration method. An explicit expression of the $8N$ mean-field equation can be, for instance, found in Appendix A of Ref. \cite{Siw25}. In practice, we solve the mean-field equations of motion using an adaptive fourth-order Runge-Kutta integration method, with $500$ time steps (we remind that the time is expressed in units of the inverse coupling constant $\frac{1}{\mu}$). 

As an illustration of the inability of the mean-field method to describe quantum many-body correlations, we compare in Figure \ref{fig:mfvsexact} the oscillation of the transition probability $P_{\nu\rightarrow \nu_X}(t)$ of neutrinos in a 8-neutrino system. In the flavor evolution $P_{\nu\rightarrow \nu_X}(t)$, the index $\nu$ denotes a given neutrino and $\nu_X$ refers to any of the three possible flavors: $\nu_{e^-}, \nu_{\mu}$ or $\nu_{\tau}$. The system is initialized in the state $| \nu_e\nu_{\mu}\nu_e \nu_{\tau} \nu_{\tau}\nu_e\nu_{\mu}\nu_{e} \rangle$. As can be seen from the figure, the mean field solution rapidly deviates from the exact one and fails to capture the main features of the oscillation probabilities of 
different neutrinos. Although mean-field methods may provide reasonable estimates of one-body observables at short times, they generally fail to reproduce the damping seen in the long-time oscillations due to the approximate treatment of the interaction between particles.    

\subsection{Initial sampling in the flavor basis 
for the phase space approach}


In the very limited number of studies conducted so far on the three-flavor neutrino oscillation problem \cite{Tur24,Che25,Spa25,Siw23,Siw25,Che25b}, the initial conditions are typically chosen to be pure Slater determinant states of the form $| \nu_{z_1}, \cdots, \nu_{z_N}\rangle$, where $\{ | \nu_{z_\alpha}\rangle \}_{\alpha=1, \cdots N}$ are flavor eigenstates, i.e. $\{ z_\alpha \} \in(e, \mu, \tau)$. To allow for a direct comparison with these earlier works, we also consider pure initial states in the present study. However, it is important to emphasize that the phase-space method is not limited to such configurations.
Its applicability extends to arbitrary initial states, including correlated states and/or thermally equilibrated states \cite{Ayi08,Yil14}. This flexibility makes the approach a powerful and versatile tool for exploring a wide range of physical scenarios.

\subsubsection{General aspect of fluctuations in the PSA}

For the initial states considered in recent applications, initial quantum
fluctuations of one-body observables are easy to compute. Let us consider a generic one-body observable, denoted by $O$ and a system described initially by a many-body density matrix $D(0)$. The mean value
and quantum fluctuations of the observable can be evaluated through 
\begin{eqnarray}
    \langle O \rangle &=& {\rm Tr} (O D(0)), ~~ 
    \sigma^2_O = \langle O^2 \rangle - \langle O \rangle^2 .\nonumber  
\end{eqnarray}

For simple initial Slater determinant states, using the Wick theorem, one can rewrite the fluctuations as:
\begin{eqnarray}
    \sigma^2_O &=& {\rm Tr} \left[O R O (1-R)\right] \nonumber 
\end{eqnarray}
where $R$ denotes the one-body density matrix and $O$ denotes now the matrix elements of the one-body operator in the single-particle basis. Denoting generically by $| i \rangle$ the canonical basis associated with the initial state, the one-body density becomes:
$R = \sum_i | i \rangle n_i \langle i |$ 
with $n_i=1$ (resp. $0$) for hole/occupied (resp. particle/unoccupied) states. 
Accordingly, we deduce:
\begin{eqnarray}
    \langle O \rangle &=& \sum_i O_{ii} n_i, ~~ \sigma^2_O = \sum_{i,j} O_{ij} n_j O_{ji} (1 - n_i) .
\end{eqnarray}

The main idea at the heart of the PSA approach is to introduce a statistical ensemble of one-body density matrices, denoted by $R^{(\lambda)}$, where $\lambda=1, \cdots, N_{\rm evt}$ labels individual events in the ensemble. Each observable $O$ is then associated with
a corresponding set of event-specific values $\{O^{(\lambda)}\}_{\lambda=1, \cdots, N_{\rm evt}}$ defined as $O^{(\lambda)} = {\rm Tr}\left[O R^{(\lambda)}\right]$.
The expectation values of the observable $O$ are now replaced by statistical averages over the ensemble, denoted by $\overline{O} $. We have, for instance:
\begin{eqnarray}
    \overline{O} &=& \frac{1}{N_{\rm evt}} \sum_\lambda O_{ij} \overline{R^{(\lambda)}_{ji}}. \label{eq:meanstat}\\
    \overline{O^2} &=& \frac{1}{N_{\rm evt}} \sum_\lambda O_{ij} O_{kl}\overline{R^{(\lambda)}_{ji}R^{(\lambda)}_{lk}}, \label{eq:flucstat} \\
    & \cdots & \nonumber 
\end{eqnarray}
From this, we can also estimate the statistical fluctuations:
\begin{eqnarray}
    \Sigma^2_Q &=&  \frac{1}{N_{\rm evt}} \sum_\lambda O_{ij} O_{kl}\overline{ \delta R^{(\lambda)}_{ji} \delta R^{(\lambda)}_{lk}}
\end{eqnarray}
where $\delta R^{(\lambda)}_{ij}  \equiv  R^{(\lambda)}_{ij}  - \overline{R^{(\lambda)}_{ij}}$. As first shown in Ref. \cite{Ayi08} and generalized in \cite{Lac13} for general quasi-particle vacuum,
the mean value and fluctuations obtained by the statistical average identify with the quantum expectation value of the initial state, provided that we have:
\begin{eqnarray}
    \overline{R^{n(\lambda)}_{ij}} &=& \delta_{ij} n_j, \label{eq:mean} \\
\overline{\delta R^{(\lambda)}_{ij}\delta R^{(\lambda)}_{kl}} 
&=& \frac{1}{2} \delta_{il} \delta_{jk} \left[ n_i (1-n_j) + n_j (1-n_i) \right] .  \label{eq:flucgen}
\end{eqnarray}   
So, for simple states such as Slater determinants with $n_i=0$ or $1$, 
we only have fluctuations in the particle-hole channels. 

It is worth noting that the present section outlines how fluctuations in the one-body density matrices can be systematically obtained at the initial time. The PSA approach preserves the statistical picture for all times, by propagating each initial condition independently using the mean-field equations of motion.

The expectation values of any observable at time $t$ are then obtained using expressions (\ref{eq:meanstat}) 
or (\ref{eq:flucstat}), where the initial density is now replaced by the evolved set of densities $R^{(\lambda)} (t)$.

\subsubsection{Initial fluctuation for the neutrino case}

In the previous section,  we outlined the general philosophy of the PSA approach 
and the prescription for the initial sampling of a generic fermionic system. We now turn to the neutrino oscillation problem, where each neutrino is labeled by $\alpha = 1,\cdots N$. We further focus on the setup commonly considered in recent studies related to quantum computing, in which the initial state is a Slater determinant of the form $| \Psi \rangle = \bigotimes_{\alpha} | \nu_{z_\alpha} \rangle$.
The reduced one-neutrino density associated to a given neutrino is
a $3 \times 3$ matrix expressed as:
\begin{eqnarray}
    R_{ij}(\alpha) = {\rm Tr}_{\bar \alpha} \left[ | \Psi \rangle \langle \Psi|
    \right], \label{eq:Ralpha}
\end{eqnarray}
where $\bar \alpha$ denote the fact that the partial trace is made on all neutrinos 
indices except $\alpha$. Here $i,j$ denotes the flavors components, i.e. $i, j \in \{ \nu_e(\alpha), \nu_\mu(\alpha), \nu_\tau(\alpha) \}$. Due to the simple nature of the initial state, the exact one-body density also verifies $R = \bigotimes_\alpha R(\alpha)$ at $t=0$. 
These properties also hold in the PSA approach and for each event $\lambda$ 
we can write the total one-body density as $R^{(\lambda)} = \bigotimes_\alpha R^{(\lambda)}(\alpha)$. The reduced density associated with the neutrino $\alpha$ 
verifies:
\begin{eqnarray}
    \overline{R^{(\lambda)(\alpha)}_{ij}} &=& \delta_{ij} n_j(\alpha),
\end{eqnarray}
where $ n_j(\alpha) = 0$ or $1$ depending on which flavor is initially occupied by the neutrino. 

With the considered initial state it is also easy to verify that different neutrinos are statistically independent, i.e.
\begin{eqnarray}
    \overline{\delta R^{(\lambda)}(\alpha) \delta R^{(\lambda)}(\beta)} = 0,
\end{eqnarray}
for $\alpha \neq \beta$. 
Finally, applying the general prescription (\ref{eq:flucgen}) in the $3\times3$ reduced space of the neutrino $\alpha$, we see that the only nonzero components are between one occupied and one unoccupied state. 
Specifically, for $i$ (resp. $j$) occupied (resp. unoccupied):
\begin{eqnarray}
\left\{
\begin{array}{l}
\displaystyle
\overline{\delta R^{(\lambda)}_{ij} (\alpha) \delta R^{(\lambda)}_{ji}(\alpha) } = 
\overline{\delta R^{(\lambda)}_{ji} (\alpha)  \delta R^{(\lambda)}_{ij}(\alpha) } = \frac{1}{2}  \\
\\
\displaystyle
\overline{\delta R^{(\lambda)}_{ij} (\alpha)  \delta R^{(\lambda)}_{ij}(\alpha) } = 
\overline{\delta R^{(\lambda)}_{ji} (\alpha)  \delta R^{(\lambda)}_{ji}(\alpha) } = 0
\end{array}
\right.,
\end{eqnarray}
which can be simulated assuming $2$ independent random Gaussian variables with mean zero and variances equal to $1/4$ (for $(i,j) \in$ particle-hole or hole-particle):
\begin{eqnarray}
    R^{(\lambda)}_{ij}(\alpha)  = \delta R^{(\lambda)}_{ij} (\alpha) &=& x^{(\lambda)}_{ij}(\alpha) + i 
    y^{(\lambda)}_{ij} (\alpha) . \label{eq:flucxy}
\end{eqnarray}
Once the initial fluctuations of the density matrix elements $R^{(\lambda)}_{ij}(\alpha)$ are determined, the corresponding fluctuations in the polarization components -- used as initial conditions for the mean-field evolution equations (see Eq. (\ref{Eq:EqOfMotion})) -- are deduced from the relation:
\begin{eqnarray*}
\tilde{P}^{(\lambda)}_m (\alpha)&=& Tr[\lambda_m 
R^{(\lambda)}] ,
\end{eqnarray*}
which follows directly from equation (\ref{eq:Q_def}). 

To illustrate how the fluctuations change depending on which flavor states are initially 
occupied, we give below the three most common examples, assuming that the states 
are ordered as $\{ |\nu_e\rangle , |\nu_\mu \rangle , |\nu_\tau \rangle\}$
\begin{itemize}

\item {\bf If the neutrino $\alpha$ is initially in $|\nu_e\rangle$: }, the matrix $R^{(\lambda)} (\alpha) $ is given by:
\begin{eqnarray}
    R^{(\lambda)} (\alpha) &=& 
    \begin{pmatrix}
    1 & \delta R^{(\lambda)}_{12} (\alpha) & \delta R^{(\lambda)}_{13} (\alpha) \\
    \delta R^{(\lambda)}_{21}(\alpha)  &0 & 0 \\
    \delta R^{(\lambda)}_{31}(\alpha)  & 0 & 0 
    \end{pmatrix} . \nonumber 
\end{eqnarray} 
This implies that $(\delta P_1(\alpha) , \delta P_2(\alpha) , \delta P_4(\alpha) , \delta P_5(\alpha) )$ are random initial variables with mean values zero and variance equal to $1$.
Note that $P_6(\alpha) $ and $P_7(\alpha) $ are non-fluctuating variables 
equal to zero. $P_3(\alpha) $ and $P_8(\alpha) $ are also non-fluctuating with $P_3 (\alpha) = 1$ and $P_8 (\alpha) = \frac{1}{\sqrt{3}}$. 

\item {\bf If  $\alpha$ is initially in  $|\nu_\mu\rangle$:}
\begin{eqnarray}
    R^{(\lambda)} (\alpha) &=& 
    \begin{pmatrix}
    0 & \delta R^{(\lambda)}_{12} (\alpha) & 0  \\
    \delta R^{(\lambda)}_{21} (\alpha) &1 & \delta R^{(\lambda)}_{23} (\alpha) \\
    0 & \delta R^{(\lambda)}_{32}(\alpha)  & 0 
    \end{pmatrix} .\nonumber 
\end{eqnarray} 
$(\delta P_1(\alpha) , \delta P_2(\alpha) , \delta P_6(\alpha) , \delta P_7(\alpha) )$ are random initial variables with mean values zero and variance equal to $1$.
$P_4(\alpha) $ and $P_5(\alpha) $ are non-fluctuating variables 
equal to zero.
We have also $P_3(\alpha)  = -1$ and $P_8(\alpha)  = \frac{1}{\sqrt{3}}$. 

\item {\bf If  $\alpha$ is initially in  $|\nu_\tau\rangle$:}
\begin{eqnarray}
    R^{(\lambda)} (\alpha) &=& 
    \begin{pmatrix}
    0 & 0 & \delta R^{(\lambda)}_{13} (\alpha) \\
    0 &0 & \delta R^{(\lambda)}_{23} (\alpha) \\
    \delta R^{(\lambda)}_{31}(\alpha)  & \delta R^{(\lambda)}_{32} (\alpha) & 1 
    \end{pmatrix} .\nonumber 
\end{eqnarray} 
$(\delta P_4(\alpha) , \delta P_5(\alpha) , \delta P_6(\alpha) , \delta P_7(\alpha) )$ are random initial variables with mean values zero and variance equal to $1$.
$P_1(\alpha) $ and $P_2(\alpha) $ are non-fluctuating variables 
equal to zero. We have $P_3(\alpha)  = 0$ and $P_8(\alpha)  = -\frac{2}{\sqrt{3}}$.

\end{itemize}

\section{Results of the PSA}

\subsection{Transition probabilities}

\begin{figure}[htbp]
\begin{center}
\includegraphics[width=\linewidth]{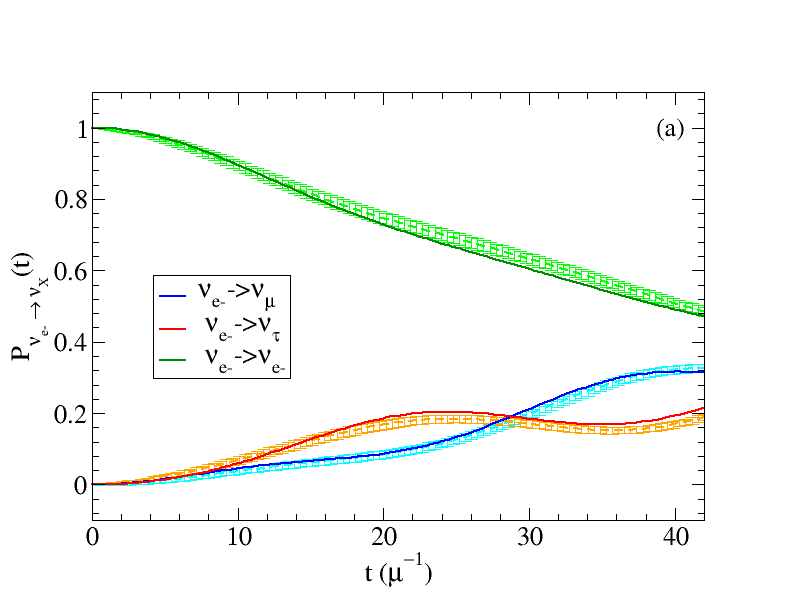}  
\includegraphics[width=\linewidth]{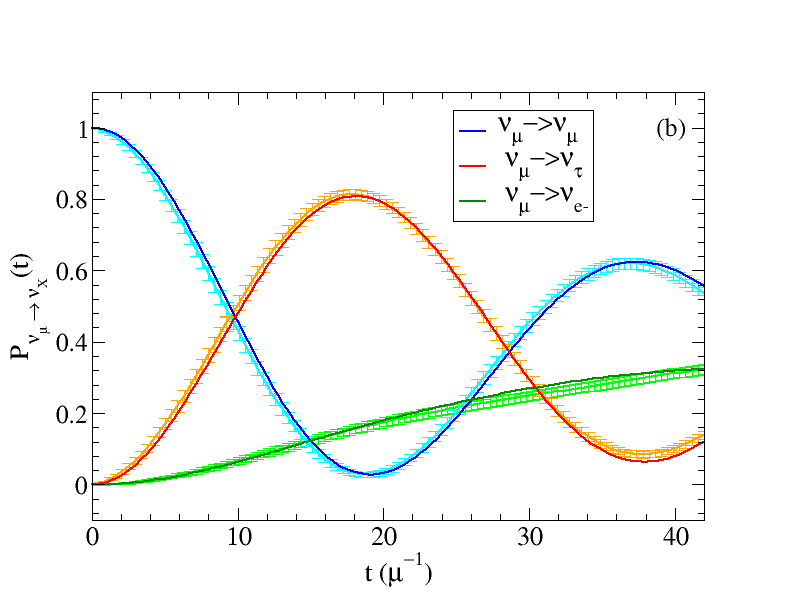}  
\includegraphics[width=\linewidth]{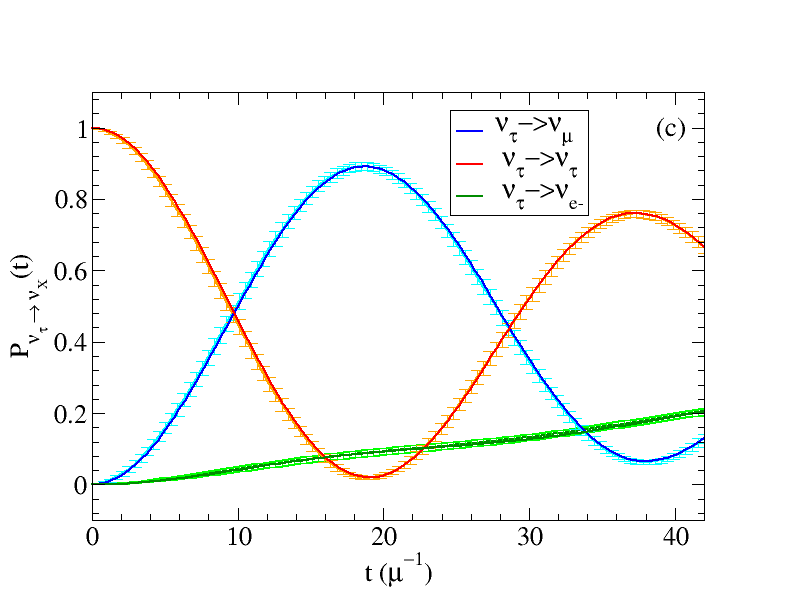}  
\caption{{\it{ Comparison of the flavor population evolutions of three selected neutrinos from a system of six interacting neutrinos 
initialized to $| \nu_e\nu_{\mu}\nu_{\tau} \nu_{\tau}\nu_{\mu}\nu_{e} \rangle$, computed with the exact  algorithm (solid line, dark colors), and PSA (dashed lines with error bars, light colors). The error bars in the phase space approach are computed using a Jackknife estimator. PSA results show clearly excellent agreement with the exact solutions. 
 }} }\label{fig:PSA_6Beams}
\end{center}
\end{figure}

\begin{figure}[htbp]
\begin{center}
\includegraphics[width=0.8\linewidth]{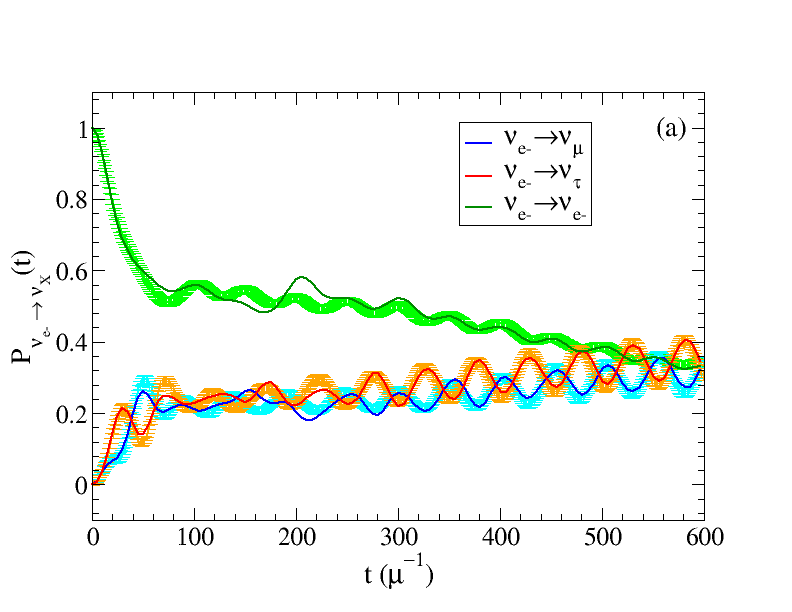}  
\includegraphics[width=0.8\linewidth]{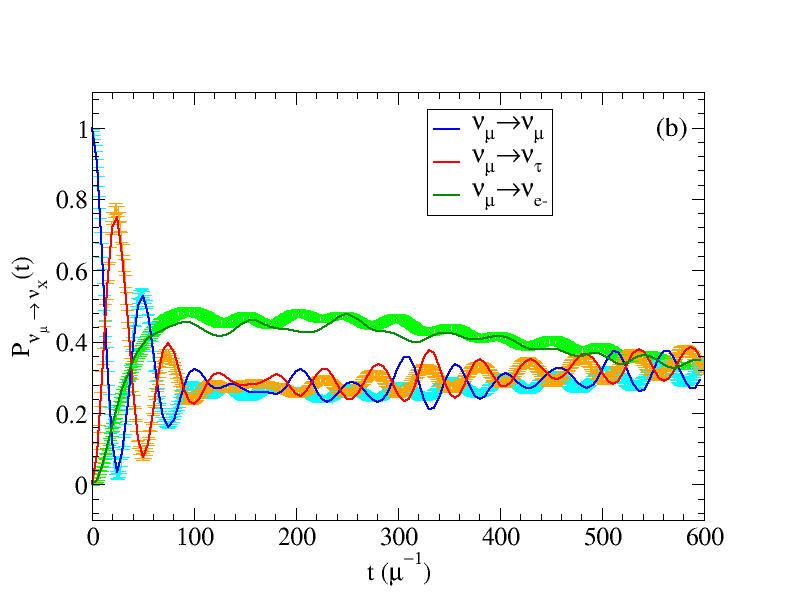}  
\includegraphics[width=0.8\linewidth]{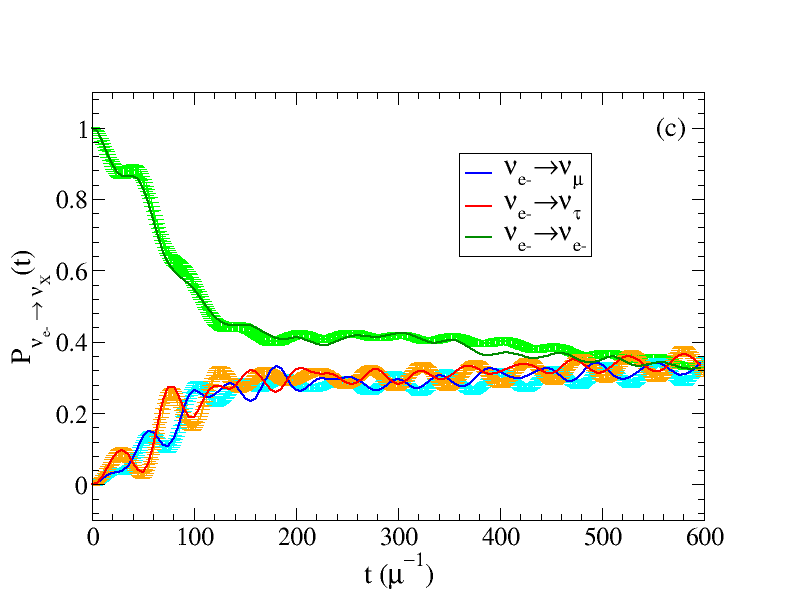}  
\includegraphics[width=0.8\linewidth]{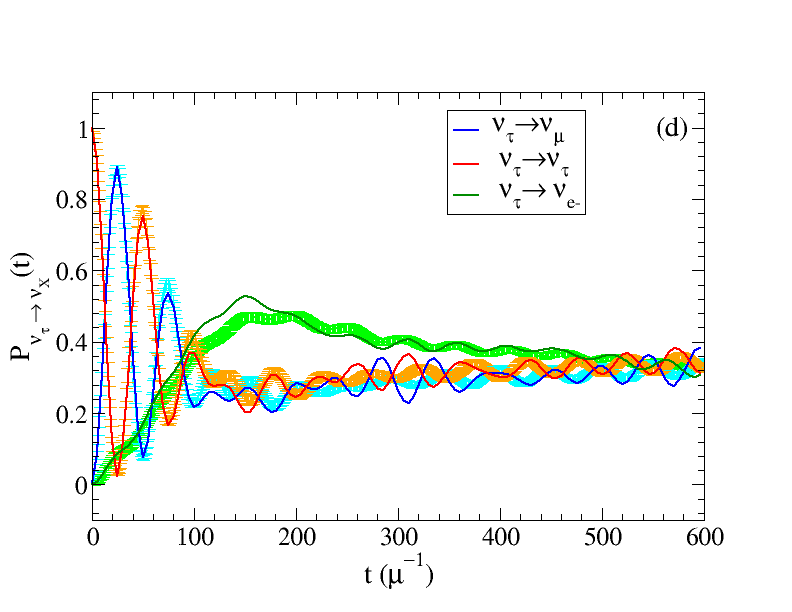} 
\caption{{\it{Comparison of the flavor population evolution of eight interacting neutrinos, initialized to $| \nu_e\nu_{\mu}\nu_e \nu_{\tau} \nu_{\tau}\nu_e\nu_{\mu}\nu_{e} \rangle$, 
computed exactly (solid lines), and with PSA (dashed lines with error bars). Considering the symmetry of the 
initial state, only the first four neutrinos are displayed in panels (a) to (d). 
PSA results and exact ones can hardly be distinguished from each other. This has to be compared with the failure of the mean-field approximation, Fig. \ref{fig:mfvsexact}. The PSA method, although based on mean-field equations, is thus much more accurate in reproducing the exact evolution, even at long times.}} 
}
\label{fig:PSA_8Beams_LongT}
\end{center}
\end{figure}

To evaluate the performance of the PSA method, we first consider systems with few neutrinos ($6$ and $8$), for which exact solutions can be computed, to compare them with PSA. In practice, the exact solution can be calculated by direct diagonalization of the Hamiltonian, or by an approximate evaluation of the evolution operator, for instance using Padé approximation. Alternatively, one can use direct propagation of the system through a quantum circuit using the Trotter-Suzuki decomposition of the Hamiltonian. In the present work, we computed the solutions both on a classical computer using the Padé approximation of the evolution operator, and on a quantum simulator, performing the evolution with the qiskit package \cite{javadi-abhari2024}.  

Since the PSA relies on a finite set of sampled trajectories, it has an inherent statistical error which decreases as the number of runs increases. 
To reliably estimate the error on the observables, we apply the so-called Jackknife binning technique \cite{Que56}. This procedure reduces the bias in the estimations of statistical errors by systematically leaving out a certain number of observations at a time and recalculating the corresponding estimates. Since the polarization is a primary quantity, the Jackknife mean coincides with the ensemble average. 

We first consider $6$ neutrinos initially in the state
$| \nu_e\nu_{\mu}\nu_{\tau} \nu_{\tau}\nu_{\mu}\nu_{e} \rangle$. This configuration is one of the benchmarks used in Ref. \cite{Tur24}. 
The exact transition from the initial flavor to different flavor populations 
are shown in Fig. \ref{fig:PSA_6Beams} and compared with the PSA results.   
The latter uses $10^4$ Metropolis iterations. The initial state being symmetric, the last 
three neutrinos (most r.h.s.) evolutions are identical to the first three (most l.h.s.) evolutions, so we plot only the results for these latter.  The flavor evolutions of these three neutrinos are shown in Figure \ref{fig:PSA_6Beams}. 
The PSA results are clearly in very good agreement with the exact solutions. 
In particular, in contrast to the standard mean-field approximation, the damping of the flavor oscillations induced by the neutrino-neutrino interaction is now properly and perfectly reproduced.   

Next, we consider a system of $8$ interacting neutrinos, initially in the state $| \nu_e\nu_{\mu}\nu_e \nu_{\tau} \nu_{\tau}\nu_e\nu_{\mu}\nu_{e} \rangle$. This state is chosen to compare with \cite{Tur24}. Figure \ref{fig:PSA_8Beams_LongT} displays the exact and the PSA calculations. For this comparison, we use a larger time interval to also explore the asymptotic behavior. 
At short times, the exact and PSA simulations are hardly distinguishable, and we observed that the PSA performs even better than in the $N=6$ case. For longer times ($t > 100 \mu^{-1}$), some deviations between the 
PSA and exact evolutions appear,  but the PSA continues to closely track the exact dynamics. At very long times, the system exhibits flavor equilibration, eventually reaching an equipartition where each flavor has a probability of $1/3$. This equilibration is again properly captured by the PSA. Note, however, that such an asymptotic behavior might not be realistic since we consider here a constant two-body interaction while in reality, when neutrinos are emitted from stellar objects, their interaction is time-dependent and gradually vanishes as the neutrinos escape from the source.

\begin{figure}[htbp]
\begin{center}
\includegraphics[height=6.5cm,width=7.8cm]{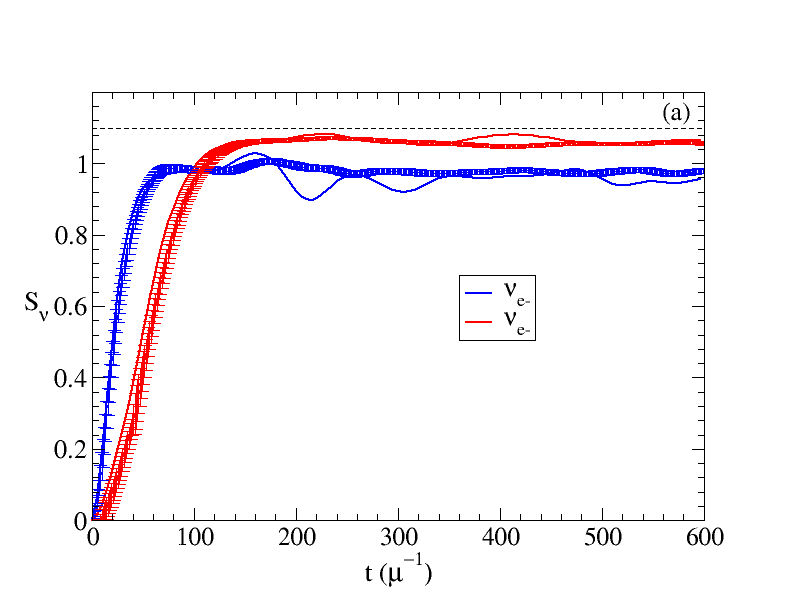}   \\
\includegraphics[height=6.5cm,width=7.8cm]{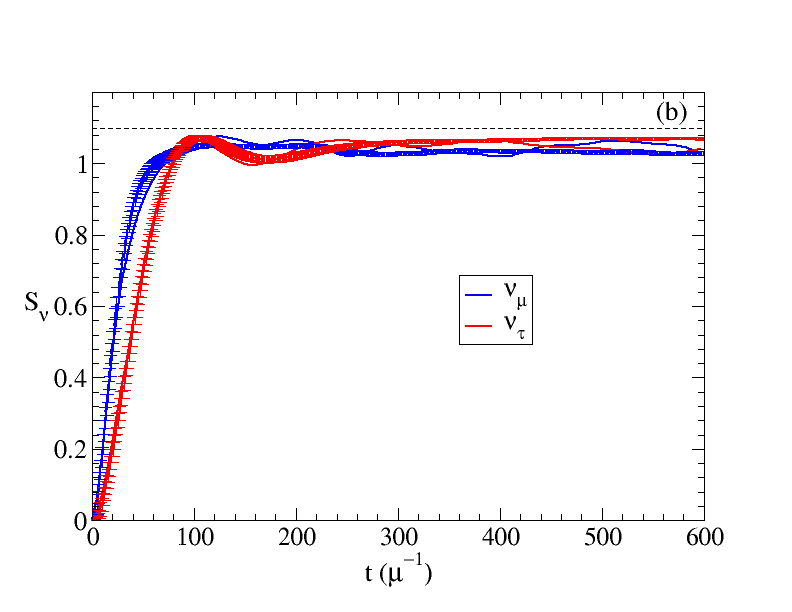}   
\caption{{\it{PSA results for single neutrino entropy (solid lines with error bars) compared to exact ones (solid lines) for $8$ neutrinos. The initial state is $| \nu_e\nu_{\mu}\nu_e \nu_{\tau} \nu_{\tau}\nu_e\nu_{\mu}\nu_{e} \rangle$. The dashed horizontal line represents the theoretical maximal value allowed for a system in the SU(3) limit.}}}
\label{Fig:PSA_N8_resultsentropy}
\end{center}
\end{figure}

\subsection{Dissipative aspects and single-neutrino entropy}

To further investigate the equilibration process, we perform a more detailed analysis of the evolution of the one-neutrino entropy. Specifically, we first discuss here how this quantity can be obtained within the PSA framework. 
For a given neutrino $\alpha$, the exact one-neutrino entropy can be evaluated using, for instance, the technique discussed in Refs. \cite{Siw23,Siw25}. 
Let $D(t)$ denote the exact many-body density matrix. From this, the expectation values of the different polarization components can be computed directly as:
\begin{eqnarray}
\vec{P}(\alpha)&=& 2 {\rm Tr} \left[ \vec{Q}(\alpha) D \right]. \label{eq:polexact}    
\end{eqnarray}
Given the eight polarization components, the exact reduced density matrix corresponding to neutrino $\alpha$ can be constructed using the following expression
\begin{eqnarray}
 R(\alpha) 
 &=& \frac{1}{3}\Big[ \mathbb{I_\alpha}+\frac{3}{2}\sum_k \lambda_k P_k(\alpha)\Big]. \label{eq:reducedexact}
\end{eqnarray}
This technique offers an alternative but equivalent method to perform the partial trace over all neutrinos except neutrino $\alpha$ in the many-body density matrix. From the reduced density matrix, we can obtain the single-neutrino entropy defined as:
\begin{eqnarray}
    S(\alpha) &=& - {\rm Tr} \left[  R(\alpha)  \ln R(\alpha) \right], \label{eq:entropyone}
\end{eqnarray}
where we adopt here the same definition of the entropy as used in Ref. \cite{Siw23}. 

The exact one-neutrino entropy for $N=8$, using the same initial conditions as in Fig. 
\ref{fig:PSA_8Beams_LongT} is shown in Fig. \ref{Fig:PSA_N8_resultsentropy}. 
In this figure, which follows the neutrino dynamics over a long timescale, we observe a typical feature of interacting systems: the single-particle entropy rises rapidly before saturating towards an asymptotic fixed value, with potential oscillations around the mean. For $N=8$, the average asymptotic is close to, but slightly below, the maximum accessible entropy value of $\ln (3) \simeq 1.1$.  

The corresponding one-neutrino entropy can also be estimated from the averaged one-neutrino entropy $\overline{R(\alpha)}$, obtained over multiple events following the same strategy.
Specifically, one can directly use Eq. \ref{eq:reducedexact}, substituting the the polarization components $\{ P_k(\alpha)\}$  with their averaged counterparts $\{ \overline{P_k(\alpha)}\}$. The results are also reported in Fig. \ref{Fig:PSA_N8_resultsentropy}, including error bars reflecting the finite number of sampled events. As expected -- given that the transition probabilities are already accurately reproduced -- the entropy growth and its saturation toward the asymptotic value are also perfectly reproduced. This confirms that the PSA framework successfully accounts for the system’s disorder and dissipation. Note that, in contrast, the mean-field theory completely fails to reproduce this behavior.

These comparisons with exact results validate the PSA method for SU(3) neutrinos. In the next section, we demonstrate that the phase space approach can be applied to systems with far more neutrinos than can be simulated exactly.

\subsection{Large-scale Phase Space simulations}

\begin{figure}[htbp]
\begin{center}
\includegraphics[width=\linewidth]{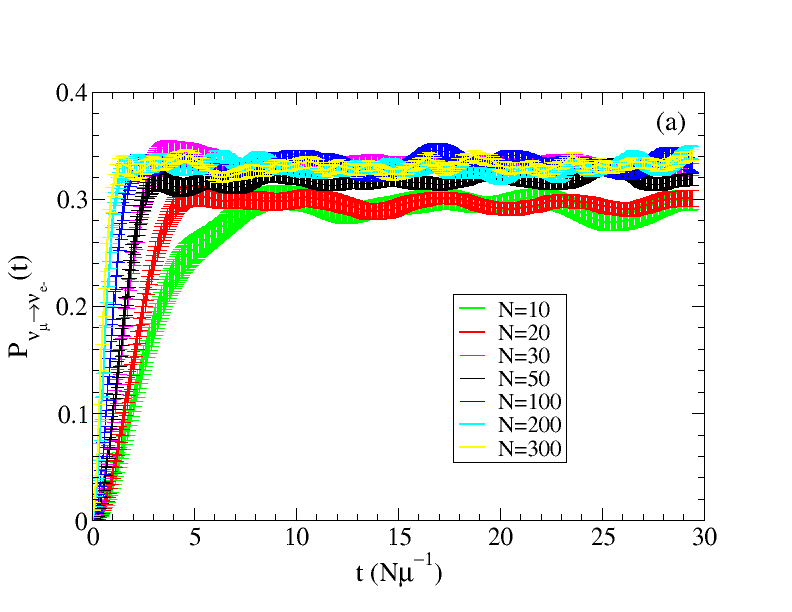} 
\includegraphics[width=\linewidth]{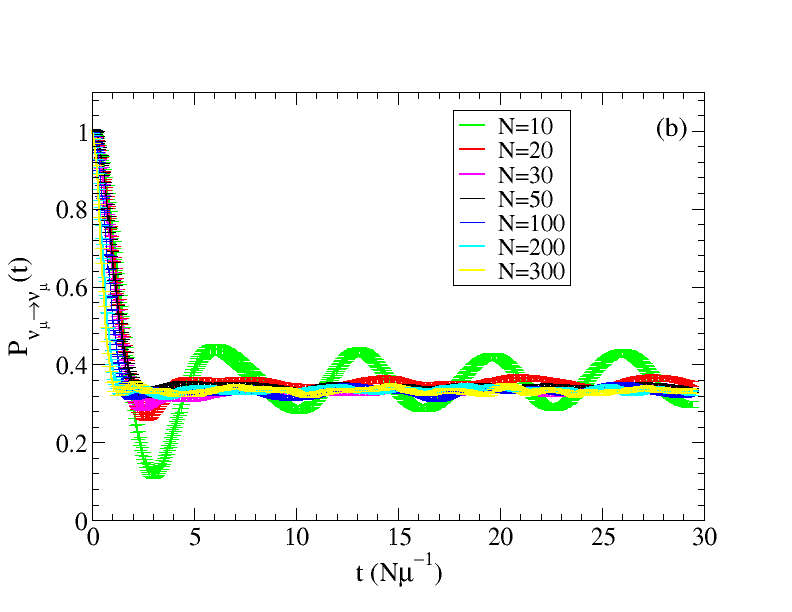} 
\includegraphics[width=\linewidth]{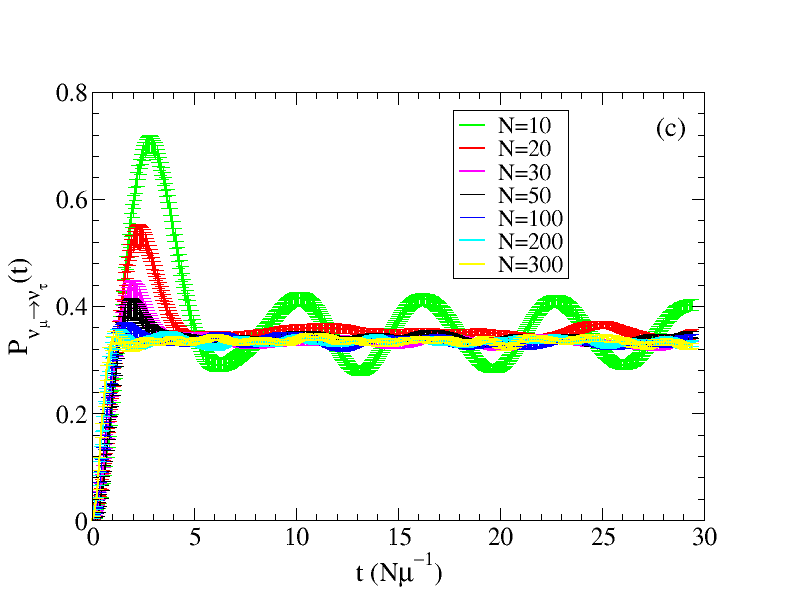} 
\caption{{\it{Oscillation probability of one neutrino embedded within a set of $N-1$ neutrinos for different total number of neutrinos $N$. The initial state is chosen as $| \nu_{\mu}\nu_{e-} \nu_{\tau}\nu_{\mu}\nu_{e-} \nu_{\tau}\ldots\rangle$.
Panels (a), (b), and (c) represent respectively the probability of the first (left most) neutrino to be in the state  
(a) $\nu_{e-}$, (b) $\nu_{\mu}$, or (c) $\nu_{\tau}$ as a function of time. In this figure, to compare different $N$, we 
have also normalized the time unit so that it does not depend on the neutrino number anymore. }}}
\label{Fig:PSA_N}
\end{center}
\end{figure}
\begin{figure}[htbp]
\begin{center}
\includegraphics[width=\linewidth]{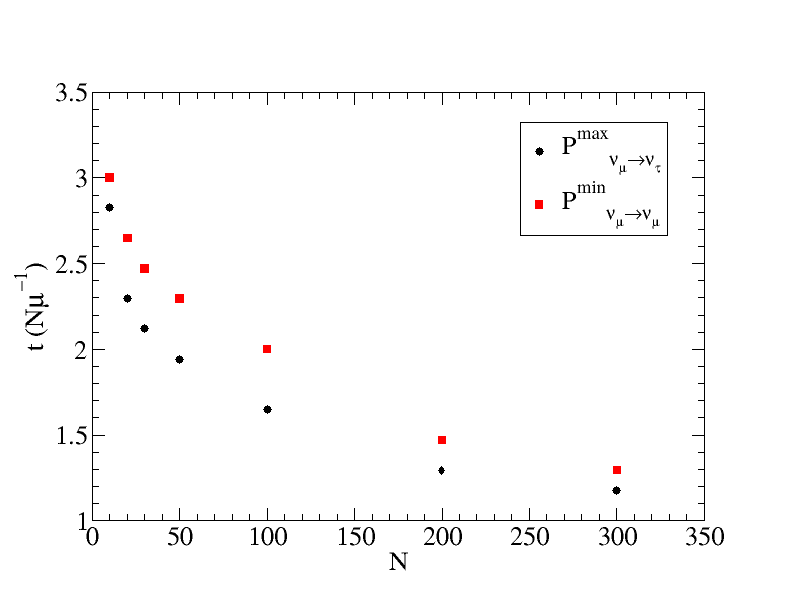} 
\caption{{\it {Normalized time at which the oscillation probability $P_{\nu_{\mu}\longrightarrow \nu_{\tau}}$
(resp. $P_{\nu_{\mu}\longrightarrow {\nu_{\mu}}}$), shown in Fig. \ref{Fig:PSA_N}
reaches its first maximum [black filled circles] (resp. minimum [red filled squares]) as a function of the total number of neutrinos. The initial state is given in the text. 
}}}
\label{Fig:Pmaxmin_N}
\end{center}
\end{figure}

In previous illustrations of the PSA approach, we focused on small systems with a number of neutrinos that does not exceed eight. In this case, the exact solutions are computable without difficulty and serve as reliable benchmarks to assess the accuracy of approximate methods. With this, we were able to demonstrate that the PSA can be efficiently extended to treat the three-flavor case, providing a highly predictive framework. One of the key advantages of the PSA lies in the fact that it is based on independent mean-field trajectories. This offers two major benefits: (i) it enables simulations of many-body systems with particle numbers well beyond the reach of exact methods ; (ii) since trajectories are independent, this approach can be straighforwardly parallelized simply by sending each trajectory to a different CPU. In this section, we showcase the scalability of the PSA approach by applying it to systems containing a large number of neutrinos, ranging from 10 up to 300.

Just to illustrate the computational challenge, consider a system of $300$ neutrinos. Representing such a system on a quantum processor using qutrits would require $300$ qutrits, corresponding to a Hilbert space having of the order of $1.37 \times 10^{143}$ states. Using a quantum computer based on qubits, the Hilbert space dimension must satisfy $2^q \ge 3^{300}$. Assuming an efficient encoding scheme for three-flavor neutrinos on qubits, this condition implies that at least $q=476$ qubits would be required to represent the system.

An interesting observable to consider is the evolution of the inversion probability of one neutrino with the total number of neutrinos $N$. In previous figures, we used the convention of Ref. \cite{Tur24}, where the time is normalized to $\mu^{-1}$ that itself scales as $N^{-1}$, which is impractical for comparing different $N$. For this reason, in the present section, we have also normalized the time unit so that it no longer depends on the total neutrino number. For the initial state, we considered a generalization of the previously discussed cases with $N=6$ and $N=8$, where each neutrino 
is initialized periodically (from left to right) within the flavors $\{ | \nu_\mu \rangle , | \nu_{e^-} \rangle , | \nu_{\tau} \rangle \}$.   

 In Fig. \ref{Fig:PSA_N} the flavor oscillation probabilities of the first neutrino (initially fixed to $\nu_{\mu}$) are displayed as a function of time, for a total number of neutrinos varying between 10 and 300. The three panels represent, respectively, the oscillation probability of the first neutrino in each of the three possible flavors: (a) $P_{\nu_{\mu}\longrightarrow \nu_{e-}}$, (b) $P_{\nu_{\mu}\longrightarrow \nu_{\mu}}$, and (c) $P_{\nu_{\mu}\longrightarrow \nu_{\tau}}$. We observe several interesting features. First, whatever the number of neutrinos $N$, the three probabilities reach asymptotically in average an equiprobable partition of the neutrino flavors with a probability $1/3$ to be in a given flavor. Second, the more neutrinos are present, the shorter is the time to reach the equipartition between flavors, and the fewer oscillations are observed around the average asymptotic 
probabilities over a long time.

\begin{figure}[htbp]
\begin{center}
\includegraphics[width=\linewidth]{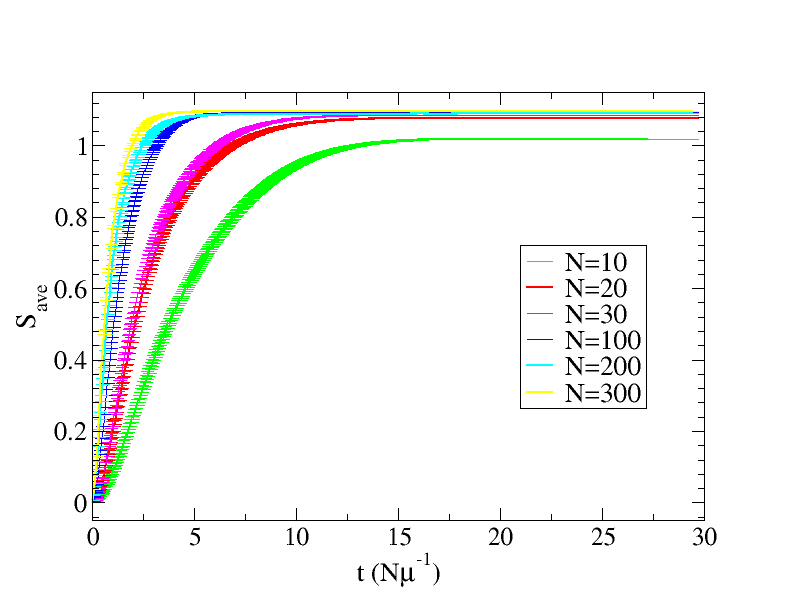} 
\caption{{\it {Average one-neutrino entropy as a function of the renormalized time for different total number of neutrinos $N$ obtained with the PSA approach. 
}}}
\label{Fig:EntropyAve_N}
\end{center}
\end{figure}

To quantify the timescale associated with the initial rise of the one-neutrino probability during the early stages of the evolution, Fig. \ref{Fig:Pmaxmin_N} presents the time 
required for the probabilities shown in Fig. \ref{Fig:PSA_N} to reach their first maximum or minimum. This figure confirms that the rising time decreases significantly
as $N$ increases. Within the considered range of $N$ shown in this figure, it is not possible to have a definitive conclusion on the scaling properties of the time needed
to increase/decrease the populations. we found that both a scaling in $\sqrt{N}$ and $\ln(N)$ can reproduce the different points shown in Fig. \ref{Fig:Pmaxmin_N}.

To complete our study, Fig. \ref{Fig:EntropyAve_N} presents the time evolution of the average one-neutrino entropy for various total numbers of neutrinos $N$.
This figure again illustrates that the entropy increases more rapidly as
$N$ increases. Moreover, the larger the value of $N$, the closer the asymptotic entropy approaches its maximum allowed value $\ln(3)$. 

Besides the entropy itself, we also analyzed the decoherence process for a single neutrino embedded in a set of neutrinos. 
Using the simplified Hamiltonian for neutrino oscillations, we clearly observed that the diagonal elements of the one-neutrino density matrix
are much larger than the off-diagonal matrix elements, especially at large time and in the large $N$ limit. Again, keeping in mind that the use of a constant two-body interaction is unrealistic, in the present case, this tends to indicate that a single neutrino embedded in a large set of neutrinos
might behave similarly to the impurity model, widely used in condensed matter \cite{Aok14}, where a subset of degrees of freedom in a lattice is
immersed into a large set of degrees of freedom that acts, and can be treated as system coupled to complex environments. Note however that the neutrino problem differs 
from condensed matter systems, due to the all-to-all coupling nature of the interaction.

\section{Conclusion}

We extend here the Phase-Space Approximation, previously applied to the two-flavor case  \cite{Lac22,Lac24} to describe neutrinos oscillating between three flavors and interacting with one another. In the small number of neutrinos regime ($N<10$), by comparing with the exact solution, we show that the PSA improves significantly upon the mean-field approximation, by properly capturing the dissipative effects arising from two-neutrino interactions. In all cases studied, PSA shows a remarkable agreement 
with the exact non-equilibrium dynamics, not only for the short-time evolution but also over longer timescales, where flavor populations equilibrate. Specifically, the timescale over which the one-neutrino entropy increases and then saturates is perfectly accounted for. 

Our overall conclusion is that the PSA method offers very high predictive power
for describing neutrino oscillations involving all three flavor components. Unlike many other approaches, the PSA has a numerical cost that scales linearly with the number of neutrinos and is naturally suited to parallelization, enabling simulations with large number of neutrinos. In this work, we present simulations involving up to 300 neutrinos. These features make PSA a highly competitive tool for studying the dynamics of neutrino systems across a wide range of sizes -- from a few to hundreds of particles. This also makes the approach a serious competitor and/or a classical computing reference for future applications of neutrino physics using quantum computers. 

Finally, we would like to stress that the PSA technique is not restricted to pure initial states as considered in this work, but can also be applied to initially correlated states or a thermal ensemble. It could also be further generalized to 
include the coupling to a surrounding environment, such as electrons, and thus allowing for the inclusion of effects like the so-called Mikhe\"iev-Smirnov-Wolfenstein (MSW) mechanism \cite{Wol78,Mik85}. We are convinced that the PSA approach, by enabling the treatment of both neutrino-neutrino and neutrino-matter interactions, together with the possibility to treat a number of neutrinos between very few to many, offers a valuable bridge between brute-force many-body simulations and more phenomenological methods (see discussions in Refs. \cite{Joh24,Kos24}).

\section{Acknowledgments }

This project has received financial support from the CNRS through the AIQI-IN2P3 project. This work is part of HQI initiative (\href{www.hqi.fr}{www.hqi.fr}) and is supported by France 2030 under the French National Research Agency award number ``ANR-22-PNQC-0002''. We also thank the QC4HEP Working Group for discussions. A. B. thanks IJCLab 
for its hospitality during his master's internship on neutrino oscillations, where part of this work was done.


\begin{thebibliography}{99}




\bibitem{Ful87} G. M. Fuller, R. W. Mayle, J. R. Wilson, and D. N.
Schramm, {\it Resonant Neutrino Oscillations and Stellar Collapse}, Astrophys. J. {\bf 322}, 795 (1987).

\bibitem{Not88} D. Notzold and G. Raffelt, {\it Neutrino dispersion at finite temperature and density}, Nucl. Phys. {\bf B 307}, 924 (1988).

\bibitem{Sig93} G. Sigl and G. Raffelt, {\it General kinetic description of relativistic mixed neutrinos},  Nucl. Phys. B 406, 423 (1993).

\bibitem{Dua06} H. Duan, G. M. Fuller, J. Carlson, and Y.-Z. Qian, {\it Simulation of coherent nonlinear neutrino flavor transformation in the supernova environment: Correlated neutrino trajectories}, Phys. Rev. {\bf D 74}, 105014 (2006).

\bibitem{Bah07}  A B Balantekin and Y Pehlivan, {\it Neutrino–neutrino interactions and flavor mixing in dense matter},  J. Phys. {\bf G 34}, 47 (2007).

\bibitem{Vol24} M. Cristina Volpe, {\it Neutrinos from dense environments: Flavor mechanisms, theoretical approaches, observations, and new directions}, 
Rev. Mod. Phys. {\bf 96}, 025004 (2024)


\bibitem{Peh11}
Y. Pehlivan, A. B. Balantekin, Toshitaka Kajino, and Takashi Yoshida, {\it Invariants of collective neutrino oscillations},  
Phys. Rev. {\bf D 84}, 065008 (2011). 


\bibitem{Bir18} Savas Birol, Y. Pehlivan, A. B. Balantekin, and T. Kajino, {\it Neutrino spectral split in the exact many-body formalism},
Phys. Rev. {\bf D 98}, 083002 (2018). 

\bibitem{Pat19} Amol V. Patwardhan, Michael J. Cervia, and A. Baha Balantekin, {\it Eigenvalues and eigenstates of the many-body collective neutrino oscillation problem}, Phys. Rev. {\bf D 99}, 123013 (2019). 


\bibitem{Rra19} Ermal Rrapaj, {\it Exact solution of multi-angle quantum many-body collective neutrino flavor oscillations}, Phys. Rev. {\bf C 101}, 065805 (2020).


\bibitem{Cer19} Michael J. Cervia, Amol V. Patwardhan, A.B. Balantekin, S.N. Coppersmith, and Calvin W. Johnson,
 {\it Entanglement and collective flavor oscillations in a dense neutrino gas}, Phys. Rev. {\bf D 100}, 083001 (2019). 
 
\bibitem{Pat21} Amol V. Patwardhan, Michael J. Cervia, and A.B. Balantekin, {\it Spectral splits and entanglement entropy in collective neutrino oscillations}, Phys. Rev. {\bf D 104}, 123035 (2021).  

\bibitem{Mar21}  Joshua D. Martin, A. Roggero, Huaiyu Duan, J. Carlson, V. Cirigliano, {\it Classical and Quantum Evolution in a Simple Coherent Neutrino Problem}, Phys. Rev. {\bf D 105}, 083020 (2022).


 \bibitem{Rog21}  Alessandro Roggero, {\it Entanglement and Many-Body effects in Collective Neutrino Oscillations }, Phys. Rev. {\bf D 104}, 103016 (2021).
  

\bibitem{Cer22}  Michael J. Cervia, Pooja Siwach, Amol V. Patwardhan, A. B. Balantekin, S. N. Coppersmith, Calvin W. Johnson, 
 {\it Collective neutrino oscillations with tensor networks using a time-dependent variational principle}, Phys. Rev. {\bf D 105}, 123025 (2022).  
 
\bibitem{Bal22}  A.B. Balantekin, {\it Quantum Entanglement and Neutrino Many-Body Systems}, J. Phys.: Conf. Ser. 2191, 012004 (2022). 

\bibitem{Xio22} Zewei Xiong, {\it Many-body effects of collective neutrino oscillations}, Phys. Rev. {\bf D 105}, 103002 (2022). 


\bibitem{Lac22} Denis Lacroix, A. B. Balantekin, Michael J. Cervia, Amol V. Patwardhan, and Pooja Siwach, {\it Role of non-Gaussian quantum fluctuations in neutrino entanglement},  Phys. Rev. {\bf D 106}, 123006 (2022). 



\bibitem{Rog22a} Alessandro Roggero, Ermal Rrapaj, and Zewei Xiong, {\it Entanglement and correlations in fast collective neutrino flavor oscillations}, 
Phys. Rev. {\bf D 106}, 043022 (2022).

\bibitem{Mar23a}  Joshua D. Martin, A. Roggero, Huaiyu Duan, J. Carlson, {\it Many-body neutrino flavor entanglement in a simple dynamic model }, arXiv:2301.07049

\bibitem{Mar23b} Joshua D. Martin, Duff Neill, A. Roggero, Huaiyu Duan, and J. Carlson, {\it Equilibration of quantum many-body fast neutrino flavor oscillations}, 
Phys. Rev. {\bf D 108}, 123010 (2023).


\bibitem{Bha23} Ramya Bhaskar, Alessandro Roggero, Martin J. Savage, {\it Time Scales in Many-Body Fast Neutrino Flavor Conversion}, Phys. Rev. {\bf C 110}, 045801 (2024). 

 
\bibitem{Lac24} Denis Lacroix, Angel Bauge, Bulent Yilmaz, Mariane Mangin-Brinet, Alessandro Roggero, A. Baha Balantekin, 
{\it Phase-Space methods for neutrino oscillations: extension to multi-beams}, Phys. Rev. {\bf D 110}, 103027 (2024). 


 

 
\bibitem{Hal21} Benjamin Hall, Alessandro Roggero, Alessandro Baroni, and Joseph Carlson, {\it Simulation of collective neutrino oscillations on a quantum computer}, Phys. Rev. {\bf D 104}, 063009. 

\bibitem{Yet22}  K\"ubra Yeter-Aydeniz, Shikha Bangar, George Siopsis, and Raphael C. Pooser, {\it Collective neutrino oscillations on a quantum computer}, Quantum Inf Process {\bf 21}, 84 (2022). 

\bibitem{Kum22} Abhishek Kumar Jha and Akshay Chatla, {\it Quantum studies of neutrinos on IBMQ processors}, Eur. Phys. J. Spec. Top. {\bf 231}, 141 (2022). 

\bibitem{Ill22a} Marc Illa, Martin J. Savage, 
{\it Multi-Neutrino Entanglement and Correlations in Dense Neutrino Systems}, Phys. Rev. Lett. {\bf 130}, 221003 (2023)

\bibitem{Ill22b} Marc Illa, Martin J. Savage, {\it Basic Elements for Simulations of Standard Model Physics with Quantum Annealers: Multigrid and Clock States}, Phys. Rev. {\bf A 106}, 052605 (2022).

\bibitem{Ami23} Valentina Amitrano, Alessandro Roggero, Piero Luchi, Francesco Turro, Luca Vespucci, Francesco Pederiva, {\it Trapped-Ion Quantum Simulation of Collective Neutrino Oscillations}, Phis. Rev. {\bf D 107}, 023007 (2023). 

\bibitem{Siw23b} P. Siwach, K. Harrison, and A. B. Balantekin, {\it Collective neutrino oscillations on a quantum computer with hybrid quantum-classical algorithm}, Phys. Rev. {\bf D 108}, 083039 (2023)


\bibitem{Tur24}  Francesco Turro, Ivan A. Chernyshev, Ramya Bhaskar, Marc Illa, {\it Qutrit and Qubit Circuits for Three-Flavor Collective Neutrino Oscillations }, Phys. Rev. {\bf D 111}, 043038 (2025).

\bibitem{Che25} I. A. Chernyshev, {\it Three-flavor collective neutrino oscillation simulations on a qubit quantum annealer}, 
Phys. Rev. {\bf D 111}, 043017 (2025).

\bibitem{Spa25} Luca Spagnoli, Noah Goss, Alessandro Roggero, Ermal Rrapaj, Michael J. Cervia, Amol V. Patwardhan, Ravi K. Naik, A. Baha Balantekin, Ed Younis, David I. Santiago, Irfan Siddiqi, Sheakha Aldaihan, {\it Collective Neutrino Oscillations in Three Flavors on Qubit and Qutrit Processors}, Phys. Rev. {\bf D 111}, 103054 (2025). 

\bibitem{Siw23}  Pooja Siwach, Anna M. Suliga, A. Baha Balantekin , {\it Entanglement in three-flavor collective neutrino oscillations}, Phys. Rev. {\bf D 107}, 023019 (2023). 

\bibitem{Siw25} Pooja Siwach, A. Baha Balantekin, Amol V. Patwardhan, Anna M. Suliga, 
{\it Exploring entanglement and spectral split correlations in three-flavor collective neutrino oscillations}, Phys. Rev. {\bf D 111}, 063038 (2025). 

\bibitem{Che25b} Ivan Chernyshev, Caroline E. P. Robin, Martin J. Savage, 
{\it Quantum Magic and Computational Complexity in the Neutrino Sector}, Phys. Rev. Research {\bf 7}, 023228 (2025). 



\bibitem{Ayi08} S. Ayik, {\it A stochastic mean-field approach for nuclear dynamics}, Phys. Lett. {\bf B 658}, 174 (2008).

\bibitem{Lac12} Denis Lacroix, Sakir Ayik, and Bulent Yilmaz, {\it Symmetry breaking and fluctuations within stochastic mean-field dynamics: Importance of initial quantum fluctuations},  Phys. Rev. {\bf C 85}, 041602(R) (2012). 

\bibitem{Lac13} Denis Lacroix, Danilo Gambacurta, and Sakir Ayik, {\it Quantal corrections to mean-field dynamics including pairing}, 
Phys. Rev. {\bf C 87}, 061302(R) (2013). 
 
\bibitem{Lac14} D. Lacroix and S. Ayik, {\it Stochastic quantum dynamics beyond mean field}, Eur. Phys. J. {\bf A 50}, 95 (2014). 


\bibitem{Yil14} Bulent Yilmaz, Denis Lacroix, and Resul Curebal, {\it Importance of realistic phase-space representations of initial quantum fluctuations using the stochastic mean-field approach for fermions}, Phys. Rev. {\bf C 90}, 054617 (2014).


\bibitem{Lac14b} Denis Lacroix, S. Hermanns, C. M. Hinz, and M. Bonitz, {\it Ultrafast dynamics of finite Hubbard clusters: A stochastic mean-field approach}, Phys. Rev. {\bf B 90}, 125112 (2014).   

\bibitem{Lac16} D. Lacroix, Y. Tanimura, S. Ayik, B. Yilmaz, {\it A simplified BBGKY hierarchy for correlated fermions from a stochastic mean-field approach} Eur. Phys. J. {\bf A 52}, 94 (2016).
 
\bibitem{Ulg19} Ibrahim Ulgen, Bulent Yilmaz, and Denis Lacroix, {\it Impact of initial fluctuations on the dissipative dynamics of interacting Fermi systems: A model case study}, Phys. Rev. {\bf C 100}, 054603 (2019). 


\bibitem{Reg18} David Regnier, Denis Lacroix, Guillaume Scamps, and Yukio Hashimoto, {\it Microscopic description of pair transfer between two superfluid Fermi systems: Combining phase-space averaging and combinatorial techniques}, 
Phys. Rev. {\bf C 97}, 034627 (2018). 


\bibitem{Czu20} Thomas Czuba, Denis Lacroix, David Regnier, Ibrahim Ulgen  and Bulent Yilmaz, 
{\it Combining phase-space and time-dependent reduced density matrix approach to describe the dynamics of interacting fermions}, 
Eur. Phys. J. {\bf A 56}, 111 (2020). 



\bibitem{Est20}   I. Esteban, M. C. González-García, M. Maltoni,
T. Schwetz, and A. Zhou,  {\it The fate of hints: updated global analysis of three-flavor neutrino oscillations}, JHEP 09, 178, arXiv:2007.14792 [hep-ph].

\bibitem{Nufit24} Nufit 5.3, www.nu-fit.org (2024).


 \bibitem{Das08} Basudeb Dasgupta and Amol Dighe, {\it Collective three-flavor oscillations of supernova neutrinos},
Phys. Rev. {\bf D 77}, 113002 (2008).


\bibitem{javadi-abhari2024}
Ali {Javadi-Abhari}, Matthew Treinish, Kevin Krsulich, Christopher~J. Wood, Jake Lishman, Julien Gacon, Simon Martiel, Paul~D. Nation, Lev~S. Bishop, Andrew~W. Cross, et~al.,
\newblock {\it Quantum computing with {{Qiskit}}}, arXiv:2405.08810.

 \bibitem{Que56} M. H. Quenouille, Biometrika, Vol. 43, No 3/4, 353 (1956).


\bibitem{Wol78} L. Wolfenstein, {\it Neutrino oscillations in matter}, Phys. Rev. {\bf D 17}, 2369 (1978). 
\bibitem{Mik85} S. P. Mikheyev et A. Yu. Smirnov, {\it Resonance amplification of oscillations in matter and spectroscopy of solar neutrinos}, Soviet Journal of Nuclear Physics, {\bf 42}, 913 (1985). 

\bibitem{Aok14} Hideo Aoki, Naoto Tsuji, Martin Eckstein, Marcus Kollar, Takashi Oka, and Philipp Werner, Rev. Mod. Phys. {\bf 86}, 779 (2014).

\bibitem{Joh24} L. Johns, {\it Neutrino many-body correlations}, Int. J. Mod. Phys. {\bf A 39}, 2450122 (2024).

\bibitem{Kos24} Anson Kost, Lucas Johns, and Huaiyu Duan, Once-in-a-lifetime encounter models for neutrino media: From coherent oscillations to flavor equilibration, Phys. Rev. {\bf D 109}, 103037 (2024).


\end{thebibliography}
\end{document}